\documentclass[prb,twocolumn,groupaddress]{revtex4}

\usepackage{graphicx}
\usepackage{tabularx}
\usepackage{capt-of}
\usepackage{color}
\usepackage{subfigure}
\usepackage{multirow}
\usepackage{array}
\usepackage{amsmath,amssymb,amsfonts,textcomp}
\usepackage{color}
\usepackage{subfigure}
\usepackage{multirow}
\usepackage{array}
\usepackage{physics}
\usepackage{booktabs}
\usepackage{array}
\newcolumntype{P}[1]{>{\centering\arraybackslash}p{#1}}
\usepackage{hyperref}

\ExplSyntaxOn
\NewDocumentCommand{\longdash}{ O{2} }
 {
  --\prg_replicate:nn { #1 - 1 } { \negthinspace -- }
 }
\ExplSyntaxOff

\newcommand{\thickhline}{%
    \noalign {\ifnum 0=`}\fi \hrule height 1pt
    \futurelet \reserved@a \@xhline
}
\begin{document}
\title{Spin-flip pair-density functional theory: A practical approach to treat
  static and dynamical correlations in large molecules}

\author{Oinam Romesh Meitei}
\affiliation{Department of Chemistry, Virginia Tech, Blacksburg, Virginia 24061, USA}

\author{Nicholas J. Mayhall}
\email{nmayhall@vt.edu}
\affiliation{Department of Chemistry, Virginia Tech, Blacksburg, Virginia 24061, USA}

\begin{abstract}
  We present a practical approach to treat static and dynamical correlation
  accurately in large multi-configurational systems. The static correlation is
  accounted for using the spin-flip approach which is well known for capturing
  static correlation accurately at low-computational expense.  
  Unlike previous approaches to add dynamical correlation to spin-flip models
  which use perturbation theory or coupled-cluster theory, we explore the
  ability to use the on-top pair-density functional theory approaches recently
  developed by Gagliardi and co-workers (JCTC, 10, 3669, 2014). 
  External relaxations are carried out in the spin-flip
  calculations though a restricted active space framework for which a
  truncation scheme for the orbitals used in the external excitation is
  presented.
  The performance of the approach is demonstrated by computing energy gaps
  between ground and excited states for diradicals, triradicals and linear
  polyacene chains ranging from naphthalene to dodecacene. Accurate results
  are obtained using the new approach for these challenging open-shell
  molecular systems.  
\end{abstract}

\maketitle
\section{Introduction}

An accurate computational description of large molecular systems with multi-configurational
characters or strongly correlated systems still remains a challenge  
due to the lack of a general approach that delivers high accuracy while remaining 
computationally affordable.

The most widely used approach for treating multi-configurational systems is the
complete active space self-consistent field (CASSCF) method in which the full
configurational interaction (CI) is solved within a chosen active
space with the orbitals and the CI coefficients optimized
\cite{casscf0,casscf_advanceschemphys}. The active space is typically selected with a
prior knowledge of the important 
orbitals contributing to the target chemical problem or the electronic
state (although recent advances are making it possible to algorithmically select active space orbitals \cite{autocas_knizia,autocas_reiher}). When supplied a well-defined active space, the method has 
been shown to reliably recover static correlation. However in
doing so, the bulk of the dynamical correlation is absent, precluding any hope for quantitative predictions.
A general strategy adopted to recover the dynamical correlation is to use the
complete active space second-order perturbation theory (CASPT2) approach.\cite{caspt2}
As dynamical correlation is well-described by perturbation theory, 
CASPT2 provides significant accuracy improvements compared to CASSCF, albeit with increased computational cost and memory requirements. The CASSCF method
itself is limited to 20 electrons in 20 orbital space with the current
state-of-the-art computational resources.\cite{casscf2020}
With a motivation to enable computation using larger active spaces, several
approaches have been developed such as
various flavors of selected configuration interaction (SCI)
approaches\cite{sci0,sc1,tpsci,sci_birgit1,sci_birgitta,sci_evange0,sci_evange2,sci_ghosh,sci_greer,sci_hoffman,sci_paterson,sci_sharma}, the density matrix renormalization group (DMRG)\cite{dmrg0,dmrg1,dmrg_chan0,dmrg_chan1,dmrg_reiher}, generalized
active-space self-consistent field (GASSCF) method\cite{gasscf}, the 2-RDM driven CASSCF
(v2RDM-CASSCF) method\cite{v2rdm0,v2rdm1} or the full CI quantum Monte Carlo
(FCIQMC) method\cite{fciqmc0,fciqmc1}.
With these reduced cost methods, larger active spaces can be used which begin to account for dynamical correlation as well, but generally a large amount of dynamical correlation is still missing. 

An alternative approach for treating multi-configurational systems is the
spin-flip (SF) method\cite{sf-krylov1,sf-krylov2,sfkrylova,sfkrylovb,sfkrylovc,sfkrylovd}. 
The SF method is a relatively simple approach which
uses a well-defined single-determinant reference on which spin-flipping
excitations are carried out to access electronic states starting from a
high-spin configuration. The SF methods have been further extended to remain
spin-pure, including external relaxations though a restricted active space
framework 
(RAS-$n$SF)\cite{casanova-ras-sf1,hgordon-ras-sf2012,ras-sf-ci-Hgordon,ras-sf-casanova-application,ras-S-sf-casanova,dynamic-corr-ci-sf-zimmerman,mayhall-rasS,mayhall-sf-noci,mayhall-quasi2pt}. Recently, we have also proposed a redox SF approach that
simultaneously accounts for both spin and spatial degeneracies which the SF
approach alone cannot handle\cite{sf-ip_ea,sf-soc}.
The SF approaches have favorable computational scaling for practical
applications to large multi-configurational systems.
In addition to allowing traditionally single-reference methods to be applied to multi-reference problems,
(e.g., coupled-cluster theory), spin-flip also reduces the dependence on user input, since the active-space is defined automatically once the reference spin state has
been chosen. 
As a result, prior knowledge of the orbitals that are important for the target
electronic state is not necessarily required.
Here, the active space is simply chosen as the singly occupied
orbitals. The external excitation space in RAS-$n$SF are simply the doubly
occupied and the virtual orbitals although knowledge of the important
contributing orbitals can be utilized as well.
Another advantage of the SF approach is the treatment of the
electronic states including the ground state on equal footing.
However, because the RAS-$n$SF approach excludes all double excitations outside of the active space, 
it can only deliver qualitative accuracy due to the lack of dynamical correlation.\cite{casanova-ras-sf1,hgordon-ras-sf2012,ras-S-sf-casanova,dynamic-corr-ci-sf-zimmerman,ras-sf-ci-Hgordon,mayhall-rasS,mayhall-quasi2pt}

An appealing approach to recover dynamical correlations for practical
applications to large molecular systems is the combination of multi-configurational
wavefunction methods with a density functional theory (DFT) based method.
Here, the idea is to use a qualitatively correct multi-configurational wavefunction to capture the
static correlation and include a DFT-based description of the dynamical correlation.
The idea was first introduced by Lie and Clementi where they demonstrated that
the DFT results can be substantially improved by adding the correlation energy
obtained from a multi-configurational wavefunction\cite{mcdft_clementi1,mcdft_clementi2}.
However, the approach does not separate the static and dynamical correlation,
thereby double counting the correlation energy.
Several approaches have emerged in this line which attempts to overcome the
deficiencies and improve over the previous works\cite{mcdft_perez,casdft_GRAFENSTEIN2,mcdft_perez1,mcdft_perez2,mcdft_salvetti1,mcdft_salvetti2,casdft-savin0,casdft-savin1,casdft-savin2,casdft_GRAFENSTEIN,casdft_GRAFENSTEIN1,casdft_yamanaka0,casdft_yamanaka1,casdft-gusarov1,casdft-gusarov2}.
Notable example include the
approach by Colle and Solvetti\cite{mcdft_salvetti1,mcdft_salvetti2}, CAS-DFT by Savin and co-workers\cite{casdft-savin0,casdft-savin1,casdft-savin2}, modified
CAS-DFT approaches by Grafenstein and 
Cremer\cite{casdft_GRAFENSTEIN,casdft_GRAFENSTEIN1}, Yamaguchi and 
co-workers\cite{casdft_yamanaka0,casdft_yamanaka1} and CAS-DFT using on-top
pair density by Gusarov and co-workers\cite{casdft-gusarov1,casdft-gusarov2}. A nice
review of the available approaches are provided in Ref \citenum{wftDFTreview}.
The main challenge in combining multi-configurational wavefunction with a
DFT-based description of the dynamical correlation is due to the double counting of
electron correlation. The multi-configurational wavefunction generally includes
some part of the dynamical correlation within the active space. The second
complication is the ``symmetry dilemma'' in KS-DFT\cite{casdft-savin1}. The spin densities of the
multi-configurational wavefunctions are not compatible with standard density
functionals for low spin states.

Recently, a new 
approach has been proposed by Gagliardi and co-workers to include DFT-based
correlation energy to a multi-configurational wavefunction 
which addresses both the double counting of correlation energy as well as the
``symmetry dilemma''.\cite{mcpdft0,mcpdft1_caspt2,mcpdft_accounts,si-pdft,mcpdft_diradicals} The method referred to as multi-configurational-pair
density functional theory (MC-PDFT) only uses the multi-configurational
wavefunction to compute the classical Coulomb and the kinetic energy while
the rest of the exchange-correlation energy and correction to the kinetic
energy is obtained from DFT using the on-top pair density functionals. The
success of the method has been demonstrated in conjunction with CASSCF\cite{mcpdft0,mcpdft1_caspt2},
GASSCF\cite{gas_pdft_pacene}, DMRG\cite{dmrg-pdft} and v2RDM-CASSCF\cite{v2rdm_pdft}.

In this work, we propose an alternate approach to combine MC-PDFT with the
RAS-$n$SF approaches and our recently developed SF-IP/EA approach. The goal is
to recover dynamical correlation energy in the SF approaches using MC-PDFT for
practical applications to large multi-configurational systems at a considerably
low computational cost. A brief description of the RAS-$n$SF approaches and
the MC-PDFT method is provided in Section \ref{sec:sf} and
\ref{sec:mcpdft}. A practical scheme is also presented to further reduce the
computational cost in the MC-PDFT calculation. Computational details are
provided in Section \ref{sec:comp}. The performance of the approach
is demonstrated by computing energy gaps between spin states of challenging
biradicals, triradicals and polyacenes ranging from naphthalene to
dodecacene. The results are presented in Section \ref{sec:piradical},
\ref{sec:benz} and \ref{sec:acene}. Finally, a summary is provided in Section
\ref{sec:summ}.

\section{Methods}

\subsection{RAS-$n$-SF and combined IP/EA approach}\label{sec:sf}
We give a short overview of the RAS-$n$-SF and
the combined IP/EA approaches. For further details of the methods, we refer the reader to
Refs. \citenum{sf-krylov1}, \citenum{sf-krylov-perspective}, and \citenum{sf-ip_ea}.

The spin-flip (SF) approach, proposed by Krylov, provides an efficient way to model
a large number of multi-configurational problems using only a single reference
determinant.
The key idea in SF approach is that while the various S$_z$ multiplets of high
spin states have identical electronic energies, states with maximum S$_z$ have a
single determinant representation, while lower S$_z$ multiplets are highly
multi-configurational. SF approach leverages this degeneracy, by optimizing the orbitals
for the single configurational high spin (S$_z$ = S) state, then uses
spin-flipping excitations to access the target S$_z$ manifold of states. Because
both high-spin and low-spin states appear in the target S$_z$ space,
the high-spin states and low-spin states are treated on an equal footing.

As an
example, the spin-degeneracy in the valence bonding orbitals
$\ket{\sigma^{2}\pi^{4}}$ with the anti-bonding $\sigma^{*}$ and $\pi^{*}$
orbitals upon bond dissociation of N$_2$ molecule can be resolved  using the
SF approach. Using a high-spin heptet state
($\ket{\sigma^1\pi^2\pi^{*2}\sigma^1}$) with $m_s=3$ as the reference state,
the ground singlet state can be accessed using a 3SF operator.

\begin{equation}
  \ket{\Psi}=\sum_{i<j<k,\bar{a}<\bar{b}<\bar{c}}c_{ijk}^{\bar{a}\bar{b}\bar{c}}\hat{a}_{ijk}^{\bar{a}\bar{b}\bar{c}}\ket{\Psi^{Reference}}
\end{equation}

Systems with spin as well as spatial degeneracies can be treated using a
combination of the spin-flip method and the electron addition (EA) or
elimination (IP) method. Unlike in the SF approach, the combined SF-IP/EA
approach works with the closest well-defined high spin state obtained either
by oxidizing or reducing the the system. While only spatial degeneracies arise in the previous example, 
if the cationic system is considered instead (N$_{2}^{+}$) both spin
and spatial degeneracy occur. In this case, the system is multi-configurational even in the high
spin state as one of the degenerate bonding ($\sigma^{2}\pi^{3}$) and anti-bonding
($\sigma^{*}\pi^{*}$) orbitals must be left unoccupied upon dissociation. The
ambiguity is resolved by simultaneously using the 2SF-IP operator on
the neutral N$_2$ molecule.

\begin{equation}
\ket{\Psi}=\sum_{ijk\bar{a}\bar{b}}c_{ijk}^{\bar{a}\bar{b}}\hat{a}_{ijk}^{\bar{a}\bar{b}}\ket{\Psi^{Reference}}
\end{equation}

The SF and IP/EA excitations are carried out only in the singly occupied
orbital space to ensure spin-pure state. External relaxation effects are taken
into account using RAS(S) by allowing the full set of singles excitations
defined by (h, p, hp)\cite{ras-S-sf-casanova,mayhall-rasS}. The approach without the external effects is
denoted by CAS-$n$SF. Overall, the
RAS-$n$-SF and the combined IP/EA approaches provides qualitatively accurate
descriptions of static correlation as demonstrated in earlier works with the
only disadvantage being the neglect of dynamical correlations.

\begin{table}
  \caption{Truncation schemes used for the RAS1 and RAS3 active space in the
    RAS(S) excitation in the natural orbital basis. The truncations are
    performed separately for the RAS1 and RAS3 space.} 
  \label{tab:trun}
  \begin{centering}
    \begin{tabular}{clp{0.6\linewidth}}
      \hline\hline
      \noalign{\vskip2mm}
      Scheme & Abbreviation & Description \tabularnewline[2mm]
      \hline
      \noalign{\vskip2mm}
      I      & RAS-SS       & A state specific approach, truncate the active
      space for each state separately. Each state has different number of
      orbitals in the active space\tabularnewline[2mm]
      
      II     & RAS-SA       & A state average (SA) approach, truncate based on
      an average density of the participating states. Each state has the same
      number of orbitals in the active space.\tabularnewline[2mm]
      
      III    & RAS-eff      & An effective SA approach, the active space is
      defined by the maximum of the number of orbitals in each state for a
      given threshold (separately for RAS1 and RAS3).\tabularnewline[2mm]      
      \hline\hline
    \end{tabular}
    \par\end{centering}
\end{table}

\subsection{Multi-configurational pair-density functional theory}\label{sec:mcpdft}
We shortly review the MC-PDFT approach which combines multi-configurational
methods with DFT-based methods without incurring double counting of
correlation energy and the symmetry dilemma in the context of KS-DFT\cite{mcpdft0,mcpdft1_caspt2,mcpdft_accounts}. By using
1- and 2-RDMs from a multi-configurational calculation, the electronic energy
in the MC-PDFT framework is given by

\begin{equation}
  E = \sum_{pq}h_{pq}D_{pq} + \frac{1}{2}\sum_{pqrs}g_{pqrs}D_{pq}D_{rs} + E_{ot}[\rho,\Pi]\label{eq:mc_pdft_eq}
\end{equation}

where $h_{pq}$ and $g_{pqrs}$ are the one- and two-electron integrals
respectively, $D_{pq}$ is the 1-RDM and $E_{ot}$ is the on-top energy with
$\rho$ and $\Pi$ being that total density and the on-top pair density
respectively. The last two terms are the classical Coulomb term and an on-top
density functional term which replaces the two-electron contribution in the
usual electronic energy expression for many-electron systems. Here, the
one-electron contribution containing the kinetic and electron-nuclear
potential energy as well as the classical electrostatic contribution are
directly taken from the multi-configurational wavefunction. The remainder
exchange and the correlation contributions are folded into the on-top pair
density functional.

The total density $\rho$ and the on-top pair density, $\Pi$ are defined in
terms of 1-RDM and 2-RDM obtained from the multi-configurational calculations
respectively as

\begin{equation}
  \rho(r) = \sum_{pq}\phi_p(r)\phi_q(r)D_{pq} \label{eq:rdm1}
\end{equation}
\begin{equation}
  \Pi(r) = \sum_{pqrs}\phi_p(r)\phi_q(r)\phi_r(r)\phi_s(r)D_{pqrs} \label{eq:rdm2}
\end{equation}

The on-top pair density functional used within the MC-PDFT formalism is simply
obtained with ``translated'' existing exchange-correlation functionals
employed in standard KS-DFT. The derivation of the ``translated'' and the
corresponding ``fully translated'' functionals are provided in Ref. \citenum{mcpdft-derivation}.

In the present RAS-$n$SF(-IP/EA)-PDFT strategies, the appropriate RDMs entering in
Equation \ref{eq:mc_pdft_eq} are obtained from the RAS-$n$SF(-IP/EA) wavefunction
presented in the previous section. Because the computation of the 2-RDM in the combined spaces of 
RAS1, RAS2, and RAS3 would prevent application to large systems,
we propose the following truncation scheme which improves efficiency while  producing only small affects on the final energies:
\begin{enumerate}
  \item Perform RAS-$n$SF(-IP/EA) using the full space for the external RAS(S)
    excitations.
  \item Diagonalize 1-RDMs of the RAS1 and RAS3 spaces separately to obtain natural orbitals for
    the RAS subspaces. The RAS-$n$SF methods are invariant with respect to orbital rotations within a 
    RAS subspace.
  \item Define a threshold for the orbital occupation number to separately
    truncate the RAS1 and RAS3 active space for the RAS(S) excitation in the
    natural orbital basis. The truncation schemes are tabulated in Table \ref{tab:trun}
  \item Using the truncated RAS1/RAS3 spaces repeat the RAS-$n$SF(-IP/EA) calculation
    to obtain the required one- and two-RDMs for the RAS-$n$SF(-IP/EA)-PDFT calculation.
\end{enumerate}

\begin{figure}
  \centering
  \includegraphics[width=0.7\linewidth]{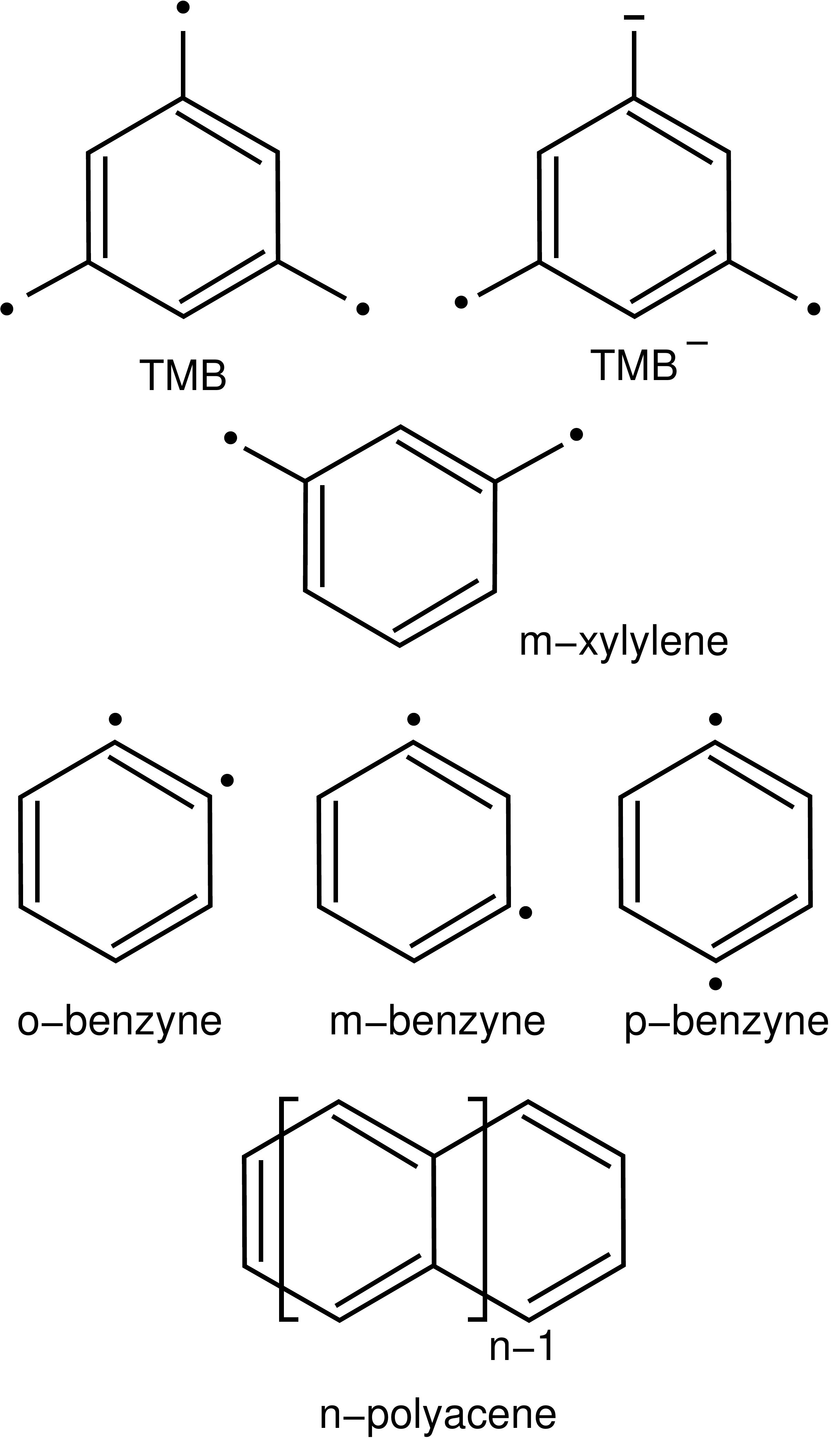}
   \caption{Schematic representations of the poly-radicals and polyacenes considered in this work.} \label{fig:mol}
\end{figure}

\section{Computational details}\label{sec:comp}

In order to determine the accuracy of the RAS-$n$SF(-IP/EA)-PDFT approach
presented in this work, we compare to existing accurate methods for the
computation of the doublet-quartet gap of triradical and singlet-triplet gaps
in aromatic biradicals and linear polyacenes and to experimental results
wherever available. The structures of 1,3,5-trimethylenebenzene (TMB) and its
negative ion (TMB$^-$) were optimized with UB3LYP/cc-pVDZ. The structure of
meta-xylylene was taken from Ref. \citenum{mx_geom_ref}, the $ortho$-, $para$- and $meta$-benzyne
radicals from Ref. \citenum{sf_krylov_benzyne}, \citenum{sf_krylov_benzyne1} and the
polyacenes from naphthalene to dodecacene from
Ref. \citenum{gas_pdft_pacene}. A schematic representation of the poly-radicals
and polyacenes considered in this work is provides in Figure \ref{fig:mol}.

The doublet state of neutral 1,3,5-trimethylenebenzene triradical was obtained
by performing a single SF operation on the quartet reference state with
$m_{s}=\tfrac{3}{2}$. The singlet and triplet state of its negative ion was
obtained by performing 1SF-EA operation on the neutral high-spin quartet
state obtained by oxidizing the target anionic state. For the benzyne radicals
as well as the polyacenes, the singlet states are obtained by performing only
a single SF operation on the triplet reference state as done in Refs. \citenum{sf_krylov_benzyne} and \citenum{casanova-ras-sf1}.

Augmented Dunning's correlation consistent aug-cc-pVDZ basis set was employed
for all the calculations\cite{ccpvdz,augccpvdz}.
The RAS-$n$SF-PDFT and the RAS-$n$SF-IP/EA-PDFT calculations were performed
with tPBE, ftPBE, tBLYP and ftBLYP on-top density functional.
Cholesky decomposition was used in all the two-electron integral calculations
with a decomposition threshold of 10$^{-6}$ a.u.\cite{molcas_cholesky}
The RAS-$n$SF(-IP/EA) and the RAS-$n$SF(IP/EA)-PDFT
calculations were performed using OpenMolcas\cite{molcas1,molcas2} and the truncation scheme described in Section \ref{sec:mcpdft} was computed using an in-house python plug-in code to OpenMolcas.

\begin{figure}[!htb] 
  \centering
  \includegraphics[width=1.0\linewidth]{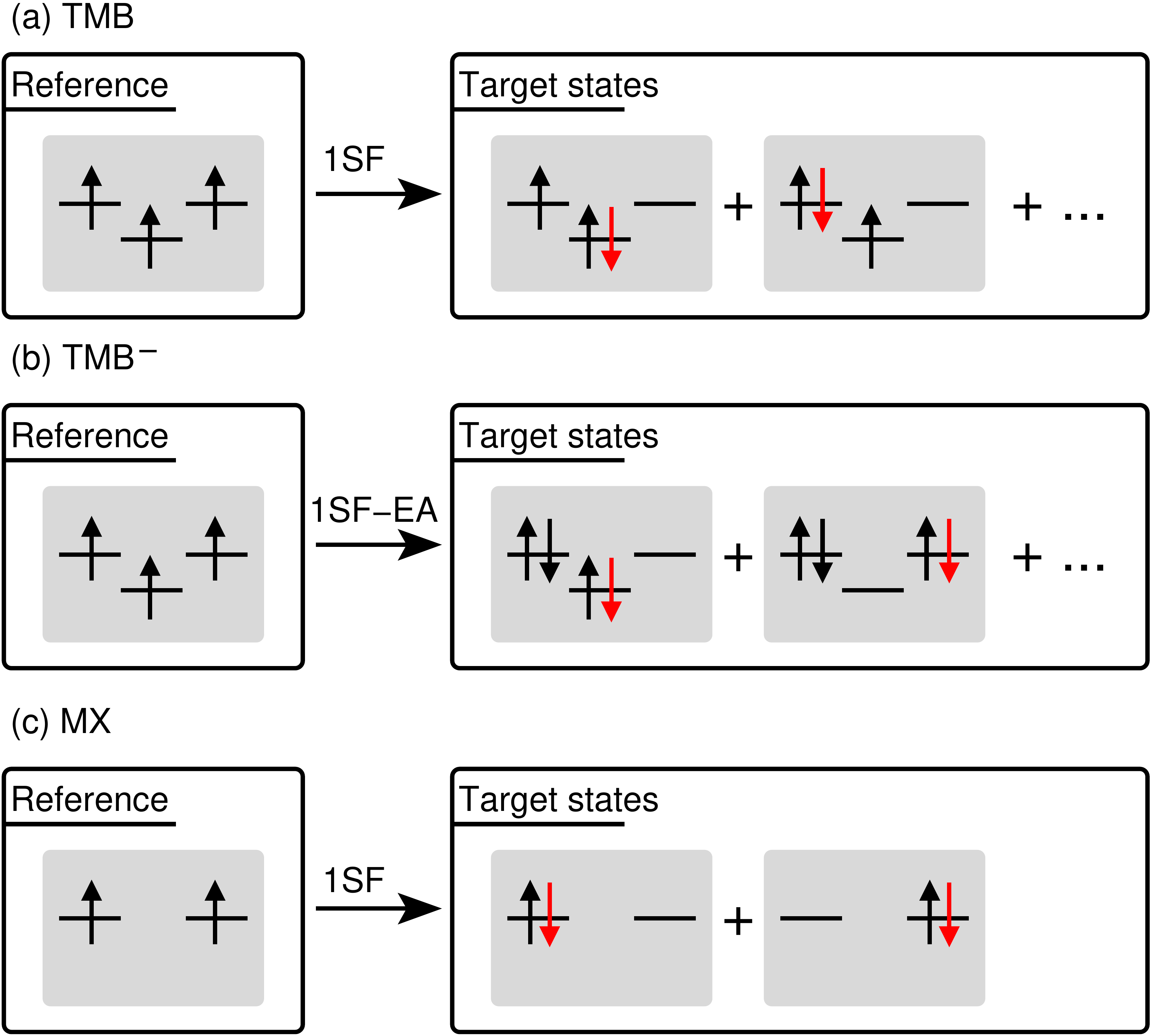}
  \caption{Illustration of the operation of 1SF and 1SF-EA on the high spin reference
    state for (a) TMB, (b) TMB$^-$ and (c) MX. The orbitals shown corresponds to the
    non-bonding orbitals and the spin-flips are indicated by
    red in color.} \label{fig:orbs}
\end{figure}

\section{Results and discussion}

\subsection{All $\pi$-polyradicals}\label{sec:piradical}

The 1,3,5-trimethylenebenezene (TMB) presents as a prototypical high-spin
all-$\pi$ triradical\cite{triradical_types}.
The triradical system has an open-shell quartet ground
state with the three unpaired electrons each occupying the nondisjoint
$\pi$ non-bonding orbitals (NBO).
Here, the three NBOs are nearly degenerate and most of the low-lying states have
heavily multiconfigurational wavefunctions.
Accurate
theoretical studies are available for the energy gap between the ground state
and the energetically lowest doublet state\cite{krylov_triradicals,tmb1,tmb_ddci,tmb3,tmb4,tmb5}. The doublet state is accesible
with just a single spin-flip operation on the high spin quartet state
($m_s=\tfrac{3}{2}$), see the sketch in Figure \ref{fig:orbs}(a).

The negative ion of TMB (TMB$^-$) on the other hand cannot be described solely by the
spin-flip approach. Four electrons occupy the nearly degenerate three NBOs and
it is ambigous to which of the three orbitals should be doubly occupied.
The ground state is a triplet state and thus is basically a
$\pi\pi$-diradical\cite{tmbanion1,tmbanion}. The ground state as well as the low-lying singlet states
can be accessed using 1SF-EA operation on the quartet reference state obtained
by oxidizing the anion. This is illustrated in Figure \ref{fig:orbs}(b).

Another prototypical all $\pi$-diradical is the $meta$-xylylene (MX) with two
electrons distributed in two nearly degenerate NBOs. The singlet-triplet gap
of MX has been intensively studied both theoretically as well as
experimentally\cite{mx_geom_ref,mayhall-rasS,mx_exp,mx3,mx2,mx1}. The diradical has an open-shell triplet ground state. The
low-lying singlet state can be accessed using a single SF operation on the
triplet reference state ($m_s=1$), see the illustration in Figure
\ref{fig:orbs}(c).

\begin{figure}[!htb] 
  \centering
  \includegraphics[width=0.8\linewidth]{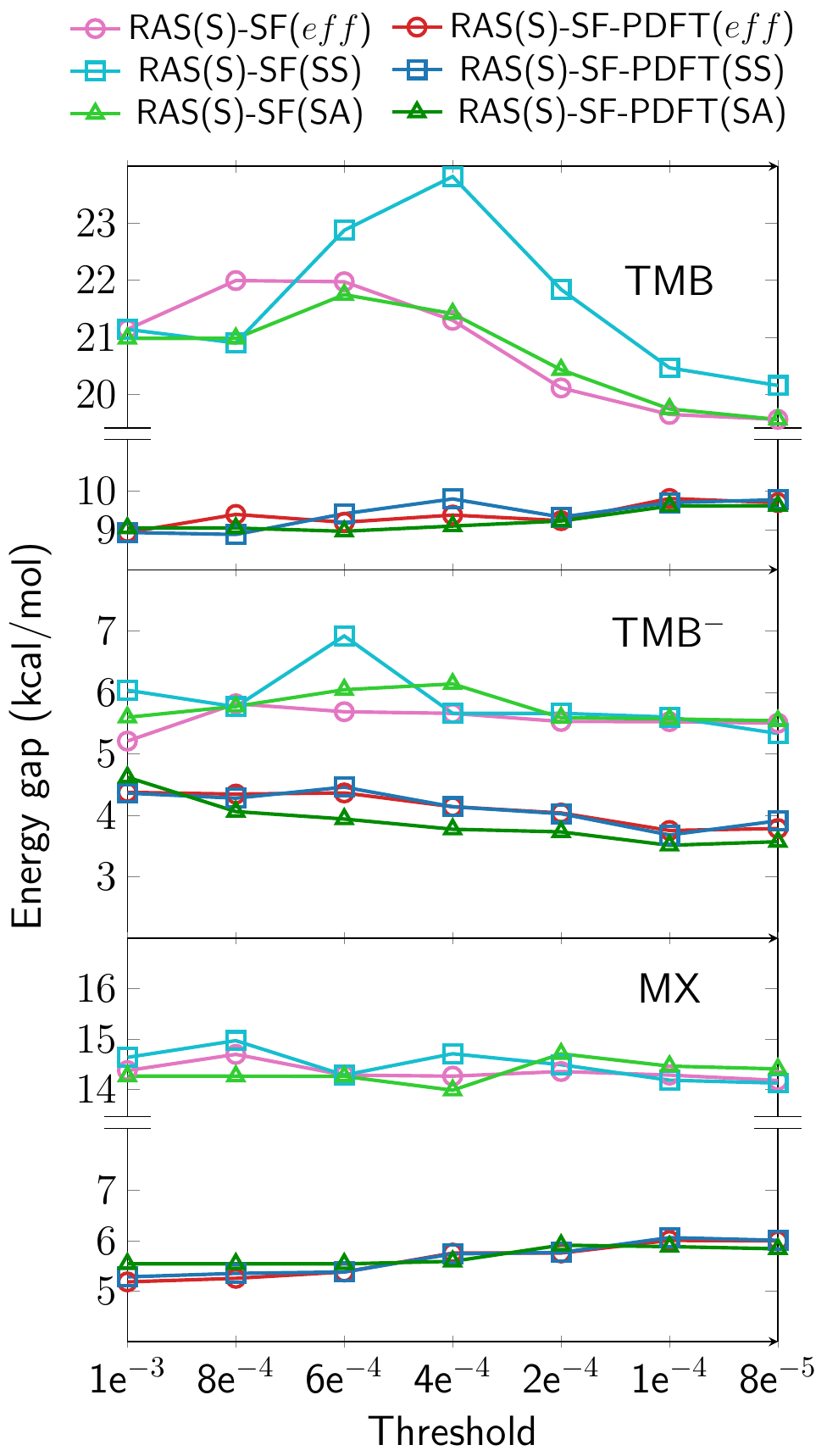}
   \caption{Convergence of energy gap w.r.t. truncation of natural orbitals,
     doublet-quartet gap of TMB (top), singlet-triplet gap of TMB$^-$ (center)
     and MX (bottom). Natural orbitals with occupation number $<$ threshold
     $or$ $> 2.0-$threshold are truncated in the RAS-$n$SF(-IP/EA)-PDFT
     computation following the truncation schemes (in parenthesis), SS, SA, \textit{eff} as outlined
     in Table \ref{tab:trun}. } \label{fig:piradical}
\end{figure}

Figure \ref{fig:piradical} plots the convergence of the different truncation schemes
described in the last part of Section \ref{sec:mcpdft}.
The natural orbitals computed from the 1-RDM of the RAS3 active space with
occupation number smaller than a given threshold are truncated in the RAS(S)
excitation scheme. The doubly occupied orbitals in the natural orbital basis
constructed from the 1-RDM of the RAS1  space is truncated with
occupation number larger than the difference of two and the given threshold.
In the three truncation schemes namely, SS, SA, and \textit{eff}, the distinction
between the SS and the \textit{eff} schemes vanishes at the convergence.
The RAS(S)-$n$SF-PDFT converges with a threshold of 0.0001 for all the
truncation schemes with the exception of SS truncation scheme in
TMB$^-$. Here, the difference in the energy gap between spin states of TMB$^-$
obtained using the SS and \textit{eff} truncation scheme is 0.13 kcal/mol at the
tightest threshold considered
which is well within the target accuracy of the presented approach.
Furthermore, the SA truncation scheme is slightly slower to converge for TMB$^-$ and MX, with the largest difference at the minimum threshold considered being 0.24 kcal/mol. Hereafter, we will consider the \textit{eff} truncation
with a threshold of 0.0001 for the RAS3 orbitals and 1.9999 for the RAS1
orbitals.

The energy gap between spin-states of TMB, TMB$^-$ and MX computed using
CAS-SF, CASSCF, RAS(S)-SF and the corresponding MC-PDFT method using the tPBE
functional is provided in Table \ref{tab:piradicals}. Literature values for
the energy gaps from previous electronic structure calculations as well the
experimental energy gap for MX are also reported in therein.
For the TMB and TMB$^-$ radicals, experimental values
were not available and so comparisons are made to the available literature
values. First, we consider the vertical doublet-quartet (DQ) energy gap 
of TMB radical. The application of MC-PDFT on CAS-1SF which does not account
for any external relaxation effect decreases the DQ energy gap by almost 15.0
kcal/mol. CAS-1SF-tPBE strongly underestimates the DQ energy gap as compared
to the previously reported gaps from CASPT2 and DDCI methods. Contrary to
this, CASSCF-tPBE increase the DQ energy gap computed with CASSCF (with
CAS(2,2)) by 7.3 kcal/mol. Interestingly, the CASSCF-tPBE energy gap is very close to a
previously reported CASSCF value with a larger active space (CAS(9,9)). On the
other hand RAS(S)-1SF-tPBE reduces the DQ energy gap from the bare RAS(S)-1SF
by almost 10 kcal/mol. Comparing to the available DDCI
and CASPT2 literature, the inclusion of dynamical correlation in the MC-PDFT
framework greatly improves the bare RAS(S)-1SF doublet-quartet energy
gap.

Next, we consider the vertical singlet-triplet (ST) gap for the anion
of TMB. Here, CAS-1SF as well as CAS-1SF-tPBE predicts a singlet ground state
as opposed to the experimental evidences for a high spin triplet ground state
for the anion\cite{tmbanion1,tmbanion}. The ST gap of bare CASSCF with
CAS(2,2) active space is increased by about 2.0 kcal/mol with the
corresponding MC-PDFT method. In contrast to this, RAS(S)-1SF-tPBE decreases
the ST gap computed using the bare RAS(S)-1SF by 1.7 kcal/mol. Here, the ST
gap agrees well with an existing CASPT2 result available in literature based
on CASSCF orbital with CAS(10,9).

For the MX diradical, the experimental ST energy gap is available and
so comparisons can be made to the adiabatic ST energy gap from the various
methods tabulated in Table \ref{tab:piradicals}. The CAS-1SF-tPBE strongly
underestimates the ST energy gap which collectively with the results obtained
for the TMB and TMB$^-$ radicals suggest that the one- and two-RDMs from the bare
CAS-1SF without any external relaxation effect is not accurate to be used in
the MC-PDFT equation. The CAS-1SF-tPBE even produces qualitatively wrong
ordering of spin-states in the TMB$^-$ radical.
On the other hand, the CASSCF-tPBE based on CAS(2,2) active space
underestimates the experimental ST gap of MX by 2.2 kcal/mol whereas the
RAS(S)-1SF-tPBE underestimates the energy gap by almost 4 kcal/mol. In
literature, the different wavefunction 
based methods tend to overestimate the ST gap of MX, see Table
\ref{tab:piradicals}. LCCQMC based on a stochastic approach from Alavi
et. al. provides the ST gap of MX very close to experimental result\cite{mxalavi}. The difficulty
in achieving an ST gap using wavefunction-based approaches comparable to the
experimental result have been demonstrated in earlier works\cite{mxcaspt2,mx2}. An extensive
study using different wavefunction methods and DFT methods for the ST gap of
MX radical can be found in Ref. \citenum{mx2}. The authors therein pointed out
the importance of including dynamical correlation with high accuracy\cite{mx2}. In our
case, while the bare RAS(S)-1SF which also includes some dynamical correlation\cite{mayhall-rasS}
overestimates the experimental ST gap by 4.3 kcal/mol, RAS(S)-1SF-tPBE
underestimates the ST by 3.9 kcal/mol.

\begin{table}
  \caption{Energy gaps between spin states of TMB\protect\footnote{Optimized geometry with UB3LYP/cc-pVDZ\label{geomtab}} (vertical doublet-quartet
    gap), TMB$^-$\textsuperscript{\ref{geomtab}} (vertical singlet-triplet
    gap) and MX\protect\footnote{Geometry from Ref. \citenum{mx_geom_ref}, UB3LYP/6-311G(d,p)} (adiabatic singlet-triplet gap) obtained
    using RAS(S)-1SF and RAS(S)-1SF-tPBE. The energy gap for TMB$^-$ was
    obtained using 1SF-EA operation. All results with aug-cc-pVDZ basis set
    and natural orbitals truncated with a threshold of 0.0001. Units are in kcal/mol.}
  \label{tab:piradicals}
  \begin{centering}
  \begin{tabular}{l@{\hspace{3mm}}P{1.5cm}P{1.5cm}P{1.5cm}}
    \hline\hline
      \noalign{\vskip2mm}
          & TMB &      TMB$^-$ &      MX \tabularnewline[2mm]
      \hline
      \noalign{\vskip2mm}
      CAS-1SF          & 17.6 & -1.9 & 4.3\tabularnewline[1mm]
      CAS-1SF-tPBE     & 2.8  & -6.1 & 1.7\tabularnewline[2mm]
      CASSCF\footnote{CAS(3,3) for TMB, CAS(4,3) for TMB$^-$ and CAS(2,2) for MX\label{castable}} &  8.3 & 2.9 &  2.1\tabularnewline[1mm]
      CASSCF-tPBE\textsuperscript{\ref{castable}}      & 15.6 & 5.0 &  7.7\tabularnewline[2mm]
      RAS(S)-1SF       & 19.6 & 5.5 & 14.2\tabularnewline[1mm]
      RAS(S)-1SF-tPBE  &  9.8 & 3.8 &  6.0\tabularnewline[2mm]
      expt.           & \longdash & \longdash & 9.6\tabularnewline[1mm]
      expt.$-$ZPE     & \longdash & \longdash & 9.9\tabularnewline[2mm]
      Literature      &
      15.7\footnote{CASSCF(9,9)/6-31G(d,p) from
          Ref. \citenum{tmb1}} ,
      11.2\footnote{DDCI/6-31G(d) using localized orbitals and
        complete virtual orbitals (fully variational) from
        Ref. \citenum{tmb_ddci}},
      13.6\footnote{CASPT2/6-31G(d,p) with CASSCF(9,9)
        orbitals from Ref. \citenum{tmb1}} &
        3.5\footnote{CASSCF(10,9)/ANO-L and CASPT2/ANO-L using
          the same CASSCF orbitals from Ref. \citenum{tmbanion1}},
          1.9\footnote{state averaged MS-CASPT2/ANO-L from Ref. \citenum{tmbanion1}}
          &
          12.9\footnote{CASSCF/6-31G(d) with CAS(8,8) from Ref. \citenum{mxcaspt2}},
          11.7\footnote{CASPT2/6-31G(d) with CASSCF orbitals with CAS(8,8) from
            Ref. \citenum{mxcaspt2}},          
          11.3\footnote{EOM-SF-CCSD/6-31G(d) from Ref. \citenum{mx3}},          
          11.8\footnote{Multireference second-order M{\o}ller-Plesset/aug-cc-pVTZ from
            Ref. \citenum{mx2}},
          9.5\footnote{LCCQMC/6-311++g(d,p) from Ref. \citenum{mxalavi}}
          \tabularnewline[2mm]
      \hline\hline
    \end{tabular}
    \par\end{centering}
\end{table}

\begin{table}
  \caption{Adiabatic singlet-triplet energy gap for $ortho$-, $meta$- and
    $para$-benzyne using the RAS-1SF and RAS-1SF-PDFT method. All results
    with aug-cc-pVDZ basis set and natural orbitals 
  truncated with a threshold of 0.0001. Units are in
    kcal/mol.} 
  \label{tab:benzyne}
  \begin{centering}
    \begin{tabular}{l@{\hspace{2mm}}P{1.7cm}P{1.7cm}P{1.7cm}}
      \hline\hline
      \noalign{\vskip2mm}
      & $o$-benzyne & $m$-benzyne & $p$-benzyne \tabularnewline[2mm]
      \hline
      \noalign{\vskip2mm}
      CAS-1SF        & 16.7 & 2.3 & 0.6 \tabularnewline[1mm]
      CAS-1SF-tPBE   & 30.2 &14.1 & 3.4 \tabularnewline[2mm]
      CASSCF\footnote{using CAS(2,2) active space\label{casben}} & 27.6 &10.1 & 1.5 \tabularnewline[1mm]
      CASSCF-tPBE\textsuperscript{\ref{casben}}    & 39.0 &23.4 & 5.4 \tabularnewline[2mm]  
      RAS(S)-1SF        & 38.0 & 22.1 & 3.8 \tabularnewline[1mm]
      RAS(S)-1SF-tPBE   & 34.9 & 20.0 & 5.8 \tabularnewline[1mm]
      RAS(S)-1SF-ftPBE  & 33.0 & 17.7 & 5.4 \tabularnewline[1mm]
      RAS(S)-1SF-tBLYP  & 35.5 & 19.0 & 6.0 \tabularnewline[1mm]
      RAS(S)-1SF-ftBLYP & 33.7 & 17.0 & 5.6 \tabularnewline[2mm]
      expt.\footnote{expt. value from
  Ref. \citenum{benzyne_expt}}             & 37.5 & 21.0 & 3.8 \tabularnewline[1mm]
      expt.$-$ZPE\footnote{ZPE correction with SF-DFT/6-311G* from
  Ref. \citenum{sf_krylov_benzyne}}
      & 38.1 & 20.0 & 3.3 \tabularnewline[2mm]
      Literature &
      35.1\footnote{CASSCF using CAS(8,8) active space from
        Ref. \citenum{cramerbenzyne}\label{cascram}},
      32.6\footnote{CASPT2 using CASSCF(8,8) orbitals from
        Ref. \citenum{cramerbenzyne}\label{cascram2}}, 
      36.8\footnote{ic-MRCCSD(T) based on CAS(2,2) from Ref. \citenum{benzyne_kohna}\label{mrcc}}
      37.3\footnote{SF-CCSD(T) from
        Ref. \citenum{benzyne_krylov_sfccsdt}\label{ccsdt}} &
      
      16.1\textsuperscript{\ref{cascram}}, 19.0\textsuperscript{\ref{cascram2}},
      19.6\textsuperscript{\ref{mrcc}},
      20.6\textsuperscript{\ref{ccsdt}} &

      3.8\textsuperscript{\ref{cascram}}, 5.8\textsuperscript{\ref{cascram2}},
      4.9\textsuperscript{\ref{mrcc}},
      4.0\textsuperscript{\ref{ccsdt}}\tabularnewline[2mm]
      \hline\hline
    \end{tabular}
    \par\end{centering}
\end{table}

\subsection{Benzyne radicals}\label{sec:benz}
The $ortho$-, $meta$- and $para$-benzyne isomers have been used as
benchmarks for new theoretical approaches and at the same time challenging
because of the strongly correlated 
biradical electrons\cite{sf_krylov_benzyne,benzyne0,benzyne1}. The benzyne
isomers have a closed-shell singlet state 
with the diradical character correlating with the distance between the
unpaired electrons. As a result, the energy gap between the ground singlet and
the lowest excited triplet state decreases following the $ortho$-, $meta$- and
$para$- sequence.

The adiabatic singlet-triplet gaps of the benzyne isomers computed using bare
CAS-1SF, CASSCF, RAS(S)-1SF and the corresponding MC-PDFT methods are presented
in Table \ref{tab:benzyne} along with the available experimental values and
selected literature values for comparison.
The RAS(S)-1SF and RAS(S)-1SF-PDFT ST energy gaps were computed using
truncated natural orbitals outlined in Section \ref{sec:mcpdft}.
A threshold of 0.0001 and 1.9999 was used for the virtual and doubly
occupied space respectively following the results from Section
\ref{sec:piradical} in the external RAS(S) excitation scheme. 
The different truncation schemes, namely, SS, SA and \textit{eff} resulted in similar
ST gaps and so only results from the \textit{eff} truncation scheme is presented. The
largest difference in the ST gaps between the different truncation schemes was
only 0.22 kcal/mol. Also presented in Table \ref{tab:benzyne} are the results
from employing different on-top functionals, viz., tPBE, tBLYP and the fully
translated variants. Overall the results obtained from using tPBE performs
better than the other functionals and so we focus our discussion only to using
the tPBE functional.

The CAS-1SF and the CASSCF methods are based on minimal active spaces, i.e.,
CAS-1SF does not account for any external relaxations while the CASSCF method
is based on CAS(2,2) active space. This is reflected in the ST gaps obtained
from CAS-1SF-tPBE as compared to the experimental results (ZPE included) for
$ortho$- and $meta$-benzyne radicals. In spite of the small active space,
CASSCF-tPBE on the other hand performs well for the $ortho$-benzyne radical,
within 1.0 kcal/mol of the experimental value whereas the ST gap is
overestimated by 3.4 kcal/mol for the $meta$-benzyne radical.
Contrary to this, RAS(S)-1SF-tPBE underestimates the experimental ST gap by
3.2 kcal/mol for $ortho$-benzyne while it exactly matches the result for the
$meta$-benzyne. Here, the ST gap of $ortho$-benzyne obtained from RAS(S)-1SF
is within 0.1 kcal/mol of the experimental value. We note that the
corresponding bare RAS(S)-1SF method accounts for some dynamical correlations
but we emphasise again that the MC-PDFT method avoids double counting of
electron correlation. Comparing to literature, we find that for the $ortho$-
and $meta$-benzyne radicals, RAS(S)-1SF-tPBE is
more accurate than CASPT2 (based on CASSCF orbitals with CAS(8,8) active
space)\cite{cramerbenzyne} which is often a method of choice for including dynamical correlations
in modelling strongly correlated systems.

The $para$-benzyne radical presents as a more challenging case and has been
the focus in comparing the ST gap obtained from various highly accurate
theoretical methods to the available experimental result\cite{cramerbenzyne,mbenzyne_kohn,benzyne_kohna,benzyne_krylov_sfccsdt,mbenzyne_evangelista,mbenzyne_kohn,mbenzyne_crawford}. The discrepency
between experimental and theoretical results have been discussed in earlier
works on the diradical\cite{mbenzyne_kohn,benzyne_kohna}. K\"ohn and co-workers have pointed out the possibility
of wrong assignments of the singlet and triplet states in the
experiment\cite{mbenzyne_kohn,benzyne_kohna}. The
ST gap for the diradical presented in Table \ref{tab:benzyne} further supports
this possiblity. The ST gap from CAS-1SF-tPBE in which the bare CAS-1SF does
not account for any external relaxation agrees well within 0.1 kcal/mol of the
experimental result in contrast to RAS(S)-1SF-tPBE.
This would mean that CAS-1SF provides a more accurate one- and two-RDMS than
the RAS(S)-1SF in the MC-PDFT equation in contrast to the results obtained for
the other radicals discussed before. Furthermore, the ST gap form RAS(S)-1SF agrees well with other
literature values obtained from CASPT2 (based on CASSCF orbitals with CAS(8,8)
active space)\cite{cramerbenzyne} and ic-MRCCSD(T)\cite{benzyne_kohna}. The ST gap also agrees well with the
CASSCF-tPBE. 

\begin{figure}[!htb] 
  \centering
  \includegraphics[width=1.0\linewidth]{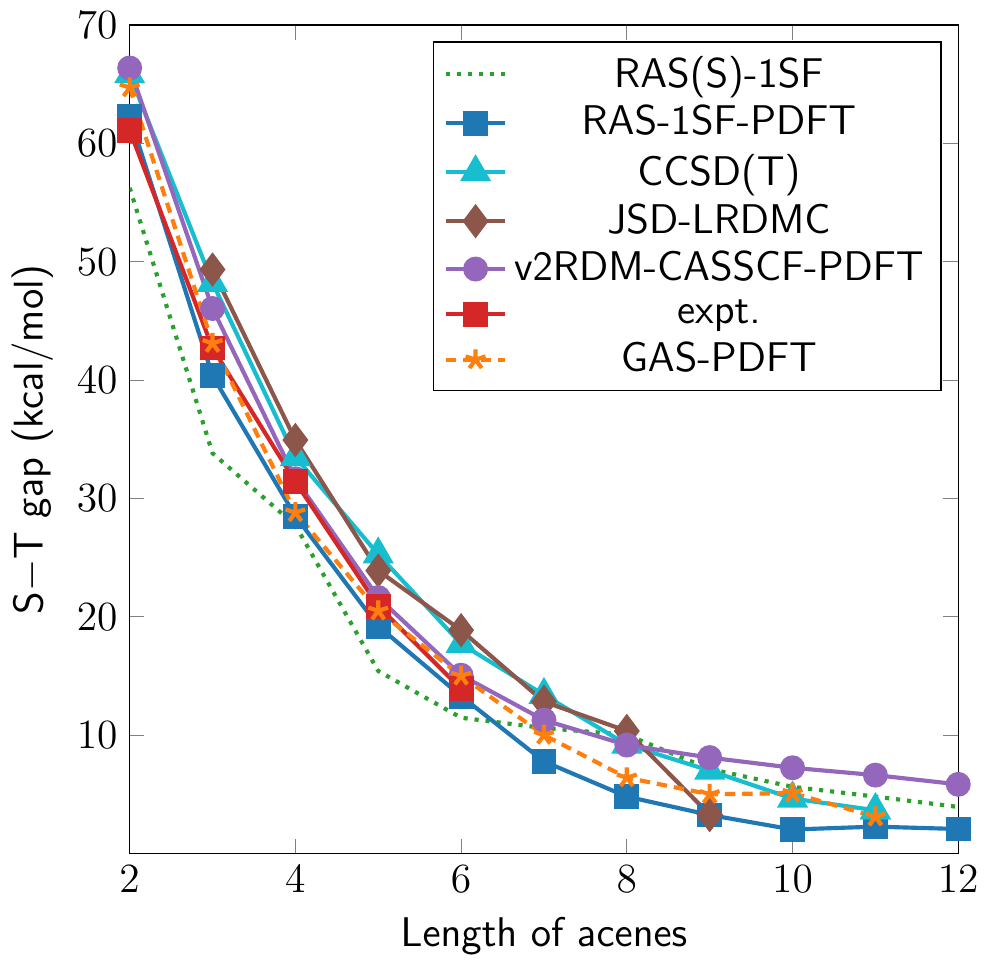}
   \caption{Adiabatic singlet-triplet gap with increasing length of
     acenes.} \label{fig:acene}
\end{figure}

\begin{table*}
  \caption{Adiabatic singlet-triplet energy gap ($E_{triplet}-E_{singlet}$)
    for polyacenes. aug-cc-pVDZ basis set are used upto Hexacene and cc-pVDZ
    basis set for the remainder. RAS(S)-1SF-PDFT results are obtained using tPBE
    functional. Units are in kcal/mol.} 
  \label{tab:acene}
  \begin{centering}
    \begin{tabular}{l@{\hspace{7mm}}c@{\hspace{3mm}}c@{\hspace{5mm}}c@{\hspace{3mm}}c@{\hspace{5mm}}c@{\hspace{3mm}}c@{\hspace{5mm}}c@{\hspace{3mm}}c@{\hspace{2mm}}}
      \hline\hline
      \noalign{\vskip2mm}
      Acene & \multicolumn{2}{c}{CAS-1SF}
      &\multicolumn{2}{c}{CASSCF\footnote{using CAS(2,2)}} &
      \multicolumn{2}{c}{RAS(S)-1SF} & expt. & expt.\tabularnewline[1mm]
      \cmidrule[0.4pt](l{2mm}r{4mm}){2-3}\cmidrule[0.4pt](l{2mm}r{2mm}){4-5}
      \cmidrule[0.4pt](l{2mm}r{2mm}){6-7}
      \noalign{\vskip1mm}
      
            &$bare$ & -PDFT   & $bare$ & -PDFT   & $bare$ & -PDFT   & & $-$ZPE\footnote{ZPE correction 
        with B3LYP/6-31G(d,p) from Ref. \citenum{dmrg-pdft}}\tabularnewline[2mm]
      \hline
      \noalign{\vskip2mm}
      Naphthalene & 52.7 & 57.4 & 64.7 & 76.8 & 56.2  & 62.3  &
      60.9\cite{acene_expt234small},   
      61.1\cite{acene_expt2big} & 64.1, 64.3
      \tabularnewline[1mm] 
      Anthracene  & 31.4 & 36.7 & 48.7 & 51.9 & 33.8 & 40.4  &
      42.7\cite{acene_expt234small},
      43.1\cite{acene_expt3big} & 45.0, 45.4
      \tabularnewline[1mm] 
      Tetracene   & 26.4 & 26.6 & 41.8 & 29.1 & 27.7 & 28.5  &
      29.5\cite{acene_expt234small} & 31.4
      \tabularnewline[1mm] 
      Pentacene   & 14.5 & 18.1 & 29.4 & 24.1 & 15.4 & 19.2  & 19.8\cite{acene_expt5} &
      20.9 \tabularnewline[1mm] 
      Hexacene    & 10.8 & 12.2 & 23.2 & 13.7 & 11.5 & 13.3  & 12.5\cite{acene_expt6} &
      13.9 \tabularnewline[1mm] 
      Heptacene   & 5.9 & 8.6 & 14.1 & 13.5 & 10.6 &  7.8 & &\tabularnewline[1mm]
      Octacene  & 3.4 & 5.6 & 10.1 & 6.4 & 10.0 & 4.8   & &\tabularnewline[1mm]
      Nonacene  & 1.3 & 4.1 & 5.0 & 6.7 &7.1 &  3.3   & &\tabularnewline[1mm]
      Decacene  & -0.2 & 1.9 & 3.2 & 2.3 & 5.6 &  2.0   & &\tabularnewline[1mm]
      Undecacene  & -1.8 & 2.6 & 0.1 & 3.4 & 4.8 &  2.3   & &\tabularnewline[1mm]
      Dodecacene  & -3.0 & 1.2 & -1.1 & 0.9 &3.9 &  2.1   & &\tabularnewline[1mm]
      \hline\hline
    \end{tabular}
    \par\end{centering}
\end{table*}

\subsection{Polyacenes} \label{sec:acene}
In the previous sections, we have shown that the RAS-$n$SF(-IP/EA)-PDFT
approach can describe challenging medium-sized biradical and triradical
molecular systems with good accuracy. In this section, we demonstrate the
performance of our approach for larger systems: the polyacenes, ranging from
naphthalene to dodecacene. The polyacenes have a singlet ground state, with
the open shell character increasing as the number of the benzene ring
increases. The singlet-triplet (ST) gap in this case correlates with the acene
length which exponentialy decreases.

Table \ref{tab:acene} reports the adiabatic singlet-triplet gaps of the 
polyacenes obtained from RAS-1SF and RAS-1SF-PDFT. Here, the on-top tPBE
functional was employed. Using the triplet state
(m$_s$=1) as the reference, the singlet state is accessed by performing only a
single SF operation on the reference state. The external RAS(S) excitations
were carried out from all the valence $\pi$ orbitals.
The adiabatic ST gaps are plotted in Figure \ref{fig:acene} along with
selected literature values, namely, CCSD(T) extrapolated according to focal
point analysis\cite{acene_ccsdt}, Monte Carlo (JSD-LRDMC)\cite{acene_motecarlo}, v2RDM-CASSCF-PDFT\cite{v2rdm_pdft} and GAS-PDFT (WFP-3
partitioning)\cite{gas_pdft_pacene}. Several other literature values are available, a detailed
analysis of the available ST gaps in literature are provided in
Ref. \citenum{acene_mr} and \citenum{acene_evangelista}

The RAS(S)-1SF-PDFT adiabatic ST gaps have a good agreement with the corresponding
experimental values as compared to the RAS(S)-1SF values.
Although the bare RAS(S)-1SF includes some dynamical correlation\cite{mayhall-rasS},
RAS(S)-1SF-PDFT improves over the bare RAS(S)-1SF method.
 
Here, a delicate balance between the dynamic and the static
correlation exist in the singlet and triplet state\cite{acene_debashree}. From Table
\ref{tab:acene}, it can be seen that the larger acenes agree very well with
the experimental value whereever available while for anthracene the difference is
$\sim$5 kcal/mol. However, we note that the experimental values were obtained
from measurements in solutions or as solids unlike the theoretical results
from gas phase calculations\cite{acene_expt234small,acene_expt2big,acene_expt3big}.
Table \ref{tab:acene} also presents ST gaps obtained from bare
CAS-1SF, CASSCF(2,2) and the corresponding MC-PDFT approaches. In both the
cases, including the PDFT correction improves upon the bare method for the ST
gaps. However, the RAS(S)-1SF-PDFT performs better in comparison to CAS-1SF-PDFT
and CASSCF-PDFT for the polyacene chains. This shows that RAS(S)-1SF provides a more accurate 1- and
2-RDMs entering the MC-PDFT equation (Equation \ref{eq:mc_pdft_eq}, \ref{eq:rdm1}
and \ref{eq:rdm2}) than CAS-1SF and CASSCF based on minal CAS(2,2) active space.

For the acenes where experimental values are not
available, the RAS-1SF-PDFT values are lower than the available literature values
shown in Figure \ref{fig:acene} although in some cases good agreements can be
seen. The CCSD(T) as well as the Monte Carlo values tend to overestimate the
available experimental values while it can be seen that the largest reported
acenes with these methods agrees well with the RAS-1SF-PDFT values.
The values reported for the
v2RDM-CASSCF-PDFT differ between 1.8 to 5.7 kcal/mol while the
difference is between 0.3 to 3.1 kcal/mol with the GAS-PDFT
method.

Another feature that can be observed in Figure \ref{fig:acene} is the
smoothly decaying exponential curve for the RAS-1SF-PDFT ST gaps which can be
fitted to the form $aexp(-bx)+c$. The fit then can be used to estimate an
extrapolated ST gap for polyacene with infinite length. Using the fitting
formula,
\begin{equation}
  E_{ST}(x)=138.14 e^{(-0.40 x)}+0.10
  \end{equation}

The adiabatic ST gap estimated for an infinitely long linear polyacene chain
is 0.10 kcal/mol.
The value is in very good agreement with the ST gap of 0.18
kcal/mol from CCSD(T)/cc-pV$\infty$Z reported in Ref. \citenum{acene_ccsdt} and with pp-RPA/cc-pVDZ with a
value between 0.0 and 2.3 kcal/mol reported in Ref. \citenum{acene-pprpa}. The later however is
estimated using vertical ST gaps.
The ST gap is
directly related to the HOMO-LUMO gap and the obtained result suggest a
closure for the HOMO-LUMO gap for inifinitly long polyacene chains.
Note that
the result is in contrast to Ref. \citenum{v2rdm_pdft} where the ST gap for
the infinite chain is reported to be 4.87 kcal/mol with the v2RDM-CASSCF-PDFT/cc-pVTZ
method and Ref. \citenum{acene_debashree} with a value of 5.06 and 5.37
kcal/mol obtained as their best estimate and from SF-CCSD/6-31+G(d,p) method respectively.
With GAS-PDFT/6-31+G(d,p) (WFP-3 partitioning), the ST gap is reported to be 1.9 kcal/mol
which is roughly half way between our value and the ones obtained from
v2RDM-CASSCF-PDFT and SF-CCSD methods\cite{gas_pdft_pacene}.
  
Although the theoretical results presented above show qualitative consistency, 
the basis sets used to compute the ST gaps differ for the various approaches and so comparison with the available literature values does not account for basis set effects.  

\section{Summary}\label{sec:summ}
In this work, we have presented a practical approach for treating large
multiconfigurational molecular system with a low computational cost.
The new method, spin-flip pair-density functional theory (SF-PDFT) uses
a spin-flip or a redox spin-flip operator to account for the static
correlation while the dynamical correlation is described with DFT using the
MC-PDFT approach.
The SF-PDFT method improves upon the result of bare spin-flip approach,
thereby capturing the missing dynamical correlations in the spin-flip
approach. In cases where the spin-flip results are already close to the
available experimental results, the SF-PDFT only changes the spin-flip results
slightly ascertaining the relaibility of SF-PDFT. The reason is because in
such cases, the static correlation dominates the electron correlation.
The method yields good accuracy for energy gaps between ground and the lowlying
excited states for challenging open-shell molecular systems. The applicability
range of the method was demonstrated by computing the singlet-triplet gap in
linear polyacene chains ranging from naphthalene to dodecacene. The SF-PDFT
predicts a closing singlet-triplet gap for an infinitly long linear polyacene
chain.

\section{Acknowledgements}
The authors are grateful for financial support provided by the U.S. Department
of Energy (Award No. DE-SC0018326) and computational infrastructure from the
Advanced Research Computing at Virginia Tech.
N. M. appreciates helpful discussions with Prof. Eugene DePrince III.

\bibliography{sfpdft.bib}
\end{document}